\documentclass[namedreferences]{solarphysics}
\usepackage[hyperref,optionalrh,solaromanenum]{spr-sola-addons} 
\usepackage{graphicx}                    
\usepackage{color}                          
\usepackage{breakurl}                     

\newcommand{\aap}{    {\it Astron. Astrophys.}}

\newcommand{\anach}{     {\it Astron. Nachrichten}}

\newcommand{\apj}{    {\it Astrophys. J.}}

\newcommand{\exas}{   {\it Exp. Astron.}}

\newcommand{\ijga}{   {\it Int. J. Geomagn. Aeron.}}
\newcommand{\jastp}{  {\it J. Atmos. Solar-Terr. Phys.}} 

\newcommand{\mnras}{  {\it Mon. Not. Roy. Astron. Soc.}}
\newcommand{\nat}{    {\it Nature}}

\newcommand{\sci}{    {\it Science}}
\newcommand{\solphys}{{\it Solar Phys.}}

\begin{document}

\begin{article}

\begin{opening}

\title{The solar rotation in the period 1853--1870 from the sunspot catalogues of Carrington, Peters, and de la Rue}
%
\author{R.~\surname{Casas}$^{1}$\sep
        J. M.~\surname{Vaquero}$^{2}$
       }

%


  \institute{$^{1}$ Institut de Ci\`encies de l'Espai (IEEC--CSIC), Campus UAB, Torre C5 parell 2n pis, E-08193 Bellaterra, Spain
                     email: \url{casas@ice.cat}.\\
             $^{2}$ Departamento de F\'\i sica, Centro Universitario de M\'erida, Universidad de Extremadura, Avda.  Santa Teresa Jornet 38, E-06800 M\'erida, Spain
                     email: \url{jvaquero@unex.es}.\\
             }


\begin{abstract}
  R. C. Carrington, C. H. F. Peters, and W. de la Rue
  observed the sunspots in the second half of the 19th
  century, determining their heliographic positions
  between 1853 and 1870, before the establishment of the
  solar program of the Royal Greenwich Observatory. The
  large tables of sunspot positions included in the
  catalogues published by these observers have recently been
  converted to a machine readable format. The present work
  analyses this data by calculating the sunspot group
  velocities for each observer. These results are then
  fitted with a differential rotation law to compare the
  data of the three observers with each other and with the
  results published by other authors.  Finally, a study is
  made of the possible relationship between the sunspot
  group areas as determined by de la Rue and the
  corresponding sunspot group velocities.
\end{abstract}

%

\end{opening}

%
\section{Introduction}
\label{intro}
The recovery of old solar observations lets one to extend
time series of various solar parameters into the past,
including solar diameter, sunspot number, sunspot areas, and
even sunspot positions. Sets of observations made during the
same time period allow gaps to be filled, and provide either
confirmation or reasonable doubt of the reliability ot those
data. In addition, with these historical data, one can see
how these parameters change with time
(\citealp{2009VaqueroVazquez}).

The telescopic observation of the Sun began in 1610 with the
records of the English astronomer Thomas Harriot, who was
followed by other pioneers. In the following decades and
centuries, until the digital era, large amounts of solar
data was collected, and is now preserved in libraries and
archives.  This data is gradually been recovered,
digitalized, and analysed with current technology
(\citealp{2009VaqueroVazquez}). In the last few years, for instance,
several software packages have been developed to analyse
sunspot drawings.  Examples are HSunspots
(\citealp{2011CristoVaqueroSanchez-Bajo}), DigiSun
(\citealp{2011Clette}), and CAMS (\citealp{2014Cakmak}).

Scans of the solar disk drawings made by the early solar
observers in the first half of the 17th century, Thomas
Harriot (\citealp{1983Shirley}), Galileo Galilei
(\citealp{1613Galileo}), Christopher Scheiner (Scheiner,
1630), and Johannes Hevelius (\citealp{1647Hevelius}), have
allowed the sunspots' heliographic positions to be determined
and, from them, the rotation velocities of the groups
(\citealp{1977EddyGilmanTrotter}; \citealp{1978Herr};
\citealp{1980Sakurai}; \citealp{1980Herr};
\citealp{1981AbarbanellWohl}; \citealp{1982Yallopetal};
\citealp{2006CasasVaqueroVazquez}). For the second half of
the 17th century, during the Maunder Minimum (1645--1715), a
period of low solar activity (\citealp{1976Eddy};
\citealp{2015VaqueroTrigo}), just some isolated observations
have been located with which to estimate the rotation velocities
of the sunspot groups. Two examples are the cases of
Nicholas Bion in 1672 (\citealp{2006CasasVaqueroVazquez})
and John Flamsteed in 1684
(\citealp{2002VaqueroSanchez-BajoGallego}).  There 
is also a long series recorded by observers at the Paris
Observatory which was recovered and analysed by
\citeauthor{1993RibesNesme-Ribes}
(\citeyear{1993RibesNesme-Ribes}).

The sunspot drawings made by Johann Staudach in the period
1749--1799 were used by \citeauthor{2012ArltFroehlich}
(\citeyear{2012ArltFroehlich}) to determine the solar
differential rotation in the second half of the 18th
century. There is only very sparse information about the solar
rotation during the first half of the 19th century,
especially during the Dalton Minimum (1800--1825).
\citeauthor{2010Sanchez-BajoVaqueroGallego}
(\citeyear{2010Sanchez-BajoVaqueroGallego}) presented an
estimate of solar rotation rate in the period 1847--1849
using the sunspot positions recorded by W. C. Bond at the
Harvard College Observatory. Moreover, \citeauthor{2011Arlt}
(\citeyear{2011Arlt}) presented an inventory of the log
books of Samuel Heinrich Schwabe, the discoverer of the
solar cycle, and an evaluation of the solar drawings
contained in this source from 1825--1867. Recently,
\citeauthor{2013Arltetal} (\citeyear{2013Arltetal})
reported the heliographic coordinates determined from those
solar drawings.

In the second half of the 19th century, long solar observation
campaigns were conducted by Richard C.
Carrington (Carrington, 1863), Gustav Sp\"orer
(\citealp{1874Sporer}), Christian F. H. Peters
(\citealp{1907Peters}), Warren de la Rue
(\citealp{1869delaRue} and \citeyear{1870delaRue}), and the
Greenwich Observatory (\citealp{2013Willisetal}).
\citeauthor{2012Lepshokov}, (\citeyear{2012Lepshokov})
reconstructed the sunspots' characteristics using the data of
Carrington and Sp\"orer. Recently,
\citeauthor{2014CasasVaquero} (\citeyear{2014CasasVaquero})
(CV14 hereafter) have described the conversion to machine
readable format of the large tables of sunspot positions
observed by Carrington, Peters and de la Rue, making a first
analysis of the datasets. We have now determined the
rotational speed of the Sun using the different sunspots as
tracers to exploit this old data and we have established a
differential rotation law for each observational dataset.
	
Our interest in this analysis was twofold. First, we wanted to
check the evolution of the differential rotation law over
short times. And second, we used the sunspot area data
published by de la Rue to search for a correlation with the
rotation velocity of the sunspot groups.

\section{Data}
\label{Data}
	
CV14 have provided an electronic version of the sunspot
position catalogues made by Carrington, Peters, and de la Rue, 
taking into consideration the different coordinate
systems that were used. Carrington and de la Rue used
normalized polar coordinates $(r, \theta)$ and the
"Carrington" heliographic coordinates for each observation.
This allowed a cross check to be made between them, revealing some
typographical errors in the original papers (see CV14 for details).  A
calculation mistake was found in the data published by
\citeauthor{1870delaRue} (\citeyear{1870delaRue}) because
the $B_0$ correction was not used in the published original
data. \citeauthor{1907Peters} (\citeyear{1907Peters})
represented the sunspot positions in celestial Cartesian
coordinates $(\Delta \alpha, \Delta \delta)$ and in the
''Peter'' heliographic coordinates. In the same paper
(\citealp{1907Peters}), a table compiled by Mr Philip Fox
helps to transform the ''Peters'' coordinates to the
''Carrington'' coordinates. This again allowed us to do a cross
check of the data and to correct some typographical errors.

A consistent dataset is required for all the data to be
analysed.  For this reason, we used the file {\bf
  comdata.dat} introduced in our previous paper (CV14). In
this file, the sunspot positions are given referred to the
Carrington heliographic coordinates calculated from the
polar and Cartesian coordinates listed by the three
historical sources, and using the Sun's physical ephemeris
provided on the {\it Horizons} Web
page\footnote{http://ssd.jpl.nasa.gov/horizons.cgi}
(\citealp{1996Giorginietal}).

We use quotes in the name of the original Carrington
coordinates because we found a systematic difference between
them and the coordinates that we calculated. We do not know the
origin of this difference, but believe it to be related
to the ephemeris (see details in CV14).

\section{Sunspot position cross-correlation}
\label{cross_correlation}
Carrington and de la Rue identified the sunspot groups with
a sequential number, trying to relate their observations
on different days. But there are some incongruences in the
data. For this reason we created an algorithm to
cross-correlate the different observations. We named this
algorithm ``Friends, and Friends of Friends'' (F\&FoF).

The basic idea is that observation $A$ is a ``friend'' of
observation $B$ if the time between them is less than 14
days, the absolute difference between their Carrington
heliographic longitudes is less than 5 degrees, and the
absolute difference between their heliographic latitudes is
less than 2 degrees.

The second rule is that if observation $A$ is a ``friend'' of
$B$, and $B$ is a ``friend'' of $C$, then observation $A$ is
a ``friend'' of $C$ too.

With this simple algorithm we found 486 different sunspots,
each with strictly more than two observations for
Carrington, 1071 for Peters, and 470 for de la Rue.

Carrington and Peters observed simultaneously from 23 May
1860 to 24 March 1861, and de la Rue's entire campaign lies
within Peters' observation run. Using the same rules as the
F\&FoF algorithm, we cross-correlated the sunspots observed
by each astronomer, finding 104 coincident groups between
Carrington (21.4\% of his total amount) and Peters (9.7\%),
and 157 between de la Rue (33.4\%) and Peters (14.6\%).

Table \ref{DiferenciesObs} lists the differences in the
heliographic coordinates between the values measured by each
pair of observers.  This table uses the median and the
standard deviation associated with the median absolute
deviation $(\sigma \approx 1.4826\cdot MAD)$ to avoid the
outliers. For both pairs of observers, the differences
between their heliographic longitude and latitude
measurements were less than $1\sigma$ in absolute value, allowing one to conclude that
the coordinates obtained are compatible.

\begin{table}
\caption{Comparison between the heliographic coordinates observed simultaneously by Carrington and Peters, and by de la Rue and Peters. The value of $\Delta$ used is the median of the differences in the observations in the sense Carrington/de la Rue minus Peters to avoid outliers, and the associated error is the standard deviation related to the MAD (see the text for the definition).}
\begin{tabular}{l r c r r c r}
Observers		&\multicolumn{3}{c}{$\Delta L$ (degrees)}&\multicolumn{3}{c}{$\Delta B$ (degrees)} \\
\hline
Carrington -- Peters	& +0.19 $\pm$ 1.55 & --0.40 $\pm$ 0.40 \\
de la Rue	-- Peters	& +0.52 $\pm$ 	1.16 & --0.05 $\pm$ 0.64 \\
\hline
\end{tabular}
\label{DiferenciesObs}
\end{table}

\section{Solar rotation velocity with the sunspots as tracers}
\label{tracers_rotation}
Once we had identified the observations of the same group
with the algorithm described in the previous section, we
determined the mean Carrington heliographic coordinates $(L,
B)$ for each group and their standard deviations $(\Delta L,
\Delta B)$.

We determined the Stonyhurst heliographic longitude
$L_{Stn}$ values for each observation using the respective Horizons
ephemeris (Giorgini {\em et al}, 1996) and the Carrington
longitudes $L_{Car}$.

To determine the synodic rotation velocity
$(\omega_{syn})$, we performed a linear regression of the
Stonyhurst longitude for each group on the observation time
$L_{Stn} = a + \omega_{syn}\cdot t$. From the fit, we also
evaluated the meridian transit time $(t_{0})$ that we used
to determine the velocity correction to apply in obtaining
the sidereal velocity $(\omega_{sid})$.

Although the dependence of the rotational velocity on the
heliographic latitude is known, we used the median and the
standard deviation associated with the MAD to reject
outliers with values beyond $3\sigma$.  For
Carrington, Peters, and de la Rue, this resulted in 15
(3.1\% of the full set of groups), 47 (4.4\%), and 60
(12.8\%) values being rejected, respectively. The large
fraction rejected of de la Rue's data is because of the  large
scatter of the sample (see CV14).

\begin{figure} 
\centerline{\includegraphics[width=1.0\textwidth,clip=]{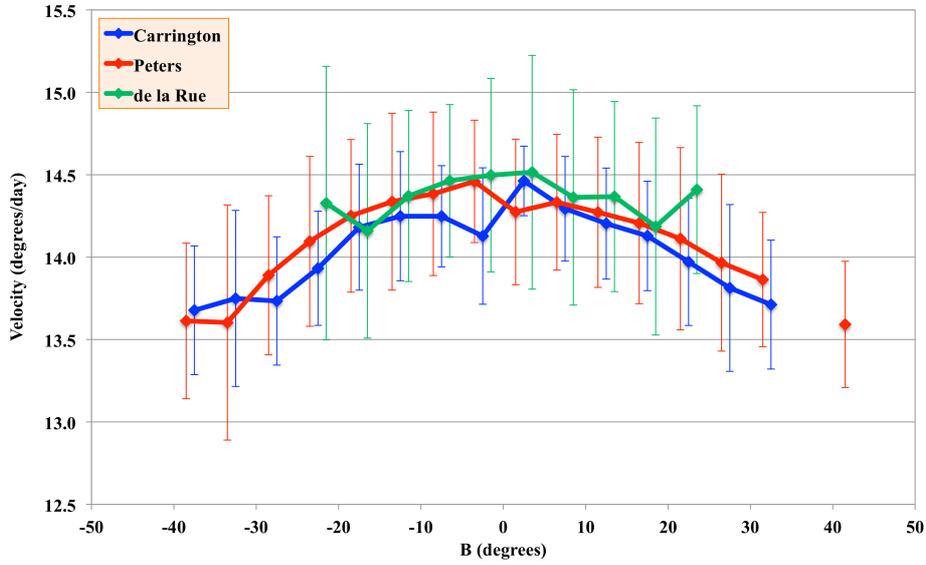}}
\caption{Averaged sidereal velocities of Carrington, Peters, and de la Rue in bins of 5 degrees latitude width. For the sake of clarity, Peters' observations are shifted $+1$ degree in latitude B, and de la Rue's are shifted $-1$ degree.}
\label{Rotation_B}
\end{figure}

The representation of the sidereal velocity as a function of
the heliographic latitude shows a major scatter. It is usual
to consider the average values of bins of latitude values.
We chose latitude bins of 5 degrees width.  Figure
\ref{Rotation_B} shows the results for the three observers.
The figure does not include the bins with only a single
measurement, and, for the sake of clarity, Peters'
heliographic latitudes are shifted $+1$ degree and those of de la
Rue $-1$ degree.

Table \ref{Vdiff_observ} lists the mean velocity differences
determined from the cross correlations between the sunspot
groups observed by two observers simultaneously.  One notes
that these values do not reflect any differences between the
rotational velocities calculated by each astronomer.

\begin{table}
\caption{Differences between the rotational velocities observed simultaneously by Carrington and Peters, and by de la Rue and Peters. The value of $\Delta\omega$ used is the median of the differences in the observations in the sense Carrington/de la Rue minus Peters' observations to avoid outliers, and the associated error is the standard deviation related to the MAD (see the text for the definition).}
\begin{tabular}{l r c r}
Observers		&\multicolumn{3}{c}{$\Delta\omega$ (degrees/day)} \\
\hline
Carrington -- Peters	& --0.11 $\pm$ 0.29 \\
de la Rue	-- Peters	& --0.03 $\pm$ 	0.45 \\
\hline
\end{tabular}
\label{Vdiff_observ}
\end{table}

\section{Differential rotation law}
\label{rotation_law} 
Various authors (\citealp{1951NewtonNunn};
\citealp{1966Ward}; \citealp{1981Tang};
\citealp{1984HowardGilmanGilman};
\citealp{1986BalthasarVazquezWohl};
\citealp{1991ZappalaZuccarello};
\citealp{2006CasasVaqueroVazquez}) have used the differential
rotation law $\omega = a + b\cdot sin^2 B$ to fit the
dependence of the rotational velocity on the heliographic
latitude when sunspots are used as tracers in different
observation sets. The term $a$ denotes the sidereal
rotation velocity for the solar equator.

Table \ref{omega_diff} lists the values we obtained using
this law.  The differential rotation laws deduced from
Peters' and de la Rue' observations are compatible at a
$1\sigma$ level. Note that the two astronomers observed
simultaneously.  While the Carrington data give an
equatorial rotation velocity that is somewhat lower, it
still lies within the $3\sigma$ interval.

\begin{table}
\caption{The observers' data fitted to an $\omega = a + b\cdot sin^2 B$ law.}
\begin{tabular}{l r c r r c r}
Observer 		&\multicolumn{3}{c}{a (degrees/day)}&\multicolumn{3}{c}{b (degrees/day)} \\
\hline
Carrington	& 14.344 $\pm$ 0.029 & --2.555 $\pm$ 0.231 \\
Peters		& 14.410 $\pm$ 0.024 & --2.233 $\pm$ 0.220 \\
de la Rue		& 14.455 $\pm$ 0.046 & --2.027 $\pm$ 0.760 \\
\hline
\end{tabular}
\label{omega_diff}
\end{table}

For comparison, the differential rotation laws obtained by
other authors are listed in Table \ref{Old_rot_data}.

\begin{table}
\caption{The differential rotation law obtained by other authors with different data sets and fitted to an $\omega = a + b\cdot sin^2 B$ law.}
\begin{tabular}{l r c r r c r}
Author/s		& \multicolumn{3}{c}{a (degrees/day)}&\multicolumn{3}{c}{b (degrees/day)} \\
Observation set &             &            &           &            &            &         \\
\hline
\citeauthor{1951NewtonNunn} (\citeyear{1951NewtonNunn})				& 14.377 &$\pm$& 0.006 & --2.77 &$\pm$& 0.08 \\
\hspace{0.5cm} GPR (1878 -- 1944)					&             &            &           &            &            &         \\
\citeauthor{1966Ward} (\citeyear{1966Ward})						& 14.523 &$\pm$& 0.006 & --2.69 &$\pm$& 0.06 \\ 
\hspace{0.5cm} GPR (1905 -- 1954)					&              &            &           &            &            &         \\
\citeauthor{1977EddyGilmanTrotter} (\citeyear{1977EddyGilmanTrotter})		& 14.37   &$\pm$& 0.04   & --0.90 &$\pm$& 0.62 \\
\hspace{0.5cm} Scheiner (1625 -- 1626)				&              &            &           &            &            &         \\
\citeauthor{1977EddyGilmanTrotter} (\citeyear{1977EddyGilmanTrotter})		& 14.94   &$\pm$& 0.12   & --4.75 &$\pm$& 3.33 \\
\hspace{0.5cm} Hevelius (1642 -- 1644)				&              &            &           &            &            &         \\
\citeauthor{1978Herr} (\citeyear{1978Herr})						& 14.58   &$\pm$& 0.12   & --2.22 &$\pm$& 1.22 \\
\hspace{0.5cm} Harriot (1611 -- 1613)				&              &            &           &            &            &         \\
\citeauthor{1981AbarbanellWohl} (\citeyear{1981AbarbanellWohl})			& 14.58   &$\pm$& 0.07   & --3.71 &$\pm$& 1.43 \\
\hspace{0.5cm} Hevelius (1642 -- 1644)				&              &            &           &            &            &         \\
\citeauthor{1982Yallopetal} (\citeyear{1982Yallopetal})					& 14.24   &$\pm$& 0.03   & --2.42 &$\pm$& 0.32 \\
\hspace{0.5cm} Scheiner (1625 -- 1626)				&              &            &           &            &            &         \\
\citeauthor{1982Yallopetal} (\citeyear{1982Yallopetal})					& 14.47   &$\pm$& 0.03   & --2.27 &$\pm$& 0.95 \\
\hspace{0.5cm} Hevelius (1642 -- 1644)				&              &            &           &            &            &         \\
\citeauthor{1984HowardGilmanGilman} (\citeyear{1984HowardGilmanGilman})		& 14.522 &$\pm$& 0.004 & --2.84 &$\pm$& 0.04 \\
\hspace{0.5cm} Mt. Wilson (1921 -- 1982)				&              &            &           &            &            &         \\
\citeauthor{1986BalthasarVazquezWohl} (\citeyear{1986BalthasarVazquezWohl})	& 14.551 &$\pm$& 0.006 & --2.87 &$\pm$& 0.06 \\
\hspace{0.5cm} GPR (1874 -- 1976)					&              &            &           &            &            &         \\
\citeauthor{1987RibesRibesBarthalot} (\citeyear{1987RibesRibesBarthalot})		& 14.22   &$\pm$& 0.07   & --8.34 &$\pm$& 0.89 \\
\hspace{0.5cm} Paris (1660 -- 1719)						&              &            &           &            &            &         \\
\citeauthor{1991ZappalaZuccarello} (\citeyear{1991ZappalaZuccarello})			& 14.643 &$\pm$& 0.015 & --2.24 &$\pm$& 0.16 \\
\hspace{0.5cm} GPR (1874 -- 1976)					&              &            &           &            &            &         \\
\citeauthor{1993Nesme-RibesFerreiraMein} (\citeyear{1993Nesme-RibesFerreiraMein})				& 14.47   &$\pm$& 0.07   & --2.00 &$\pm$& 0.31 \\
\hspace{0.5cm} Meudon (1974 -- 1984)				&              &            &           &            &            &         \\
\citeauthor{1993Nesme-RibesFerreiraMein} (\citeyear{1993Nesme-RibesFerreiraMein})				& 14.49   &$\pm$& 0.02   & --2.60 &$\pm$& 0.15 \\
\hspace{0.5cm} Mt. Wilson (1974 -- 1984)				&              &            &           &            &            &         \\
\citeauthor{1999GuptaSivaramanHoward} (\citeyear{1999GuptaSivaramanHoward})	& 14.456 &$\pm$& 0.002 & --2.89 &$\pm$& 0.02 \\
\hspace{0.5cm} Kodaikanal (1906 -- 1987)			&              &            &           &            &            &         \\
\citeauthor{2006CasasVaqueroVazquez} (\citeyear{2006CasasVaqueroVazquez})	& 14.42   &$\pm$& 0.11   & --4.96 &$\pm$& 1.40 \\
\hspace{0.5cm} Galileo (1612)						&              &            &           &            &            &         \\
\hline
\end{tabular}
\label{Old_rot_data}
\end{table}

\section{De la Rue sunspot areas and rotational velocities}
\label{areas} 
Table II of \citeauthor{1869delaRue}
(\citeyear{1869delaRue}, \citeyear{1870delaRue}) lists the
penumbral, umbral, and total areas measured for each sunspot
group. Those data are probably the first published values of
these quantities. Note that \citeauthor{2005Vaqueroetal}
(\citeyear{2005Vaqueroetal}) determined average values of the
sunspot group penumbral-to-umbral area ratio from these
early observations by de la Rue from the years 1862 to 1866 that were
similar to the values reported by Hathaway (2013) for the
period 1874--1976 using the sunspot catalogue of the Royal
Observatory, Greenwich.

\begin{figure} 
\centerline{\includegraphics[width=1.0\textwidth,clip=]{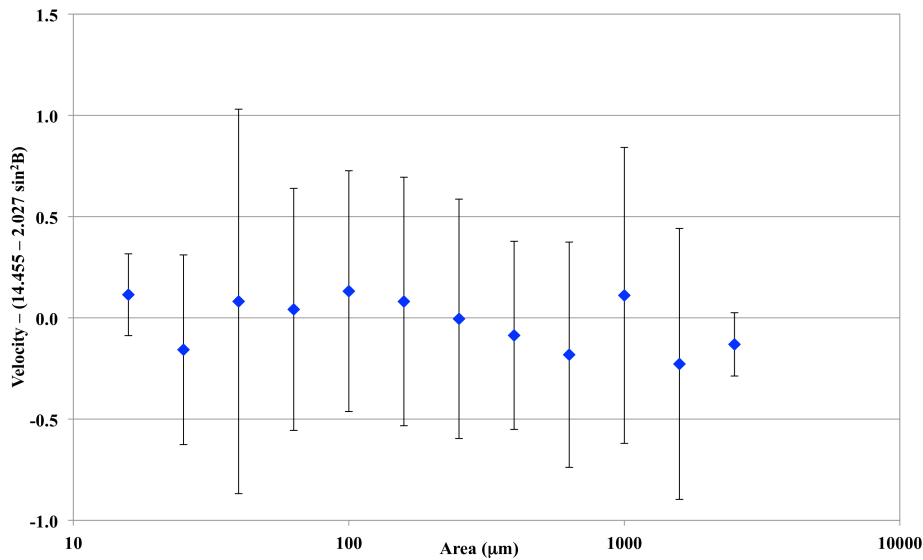}}
\caption{Averaged sunspot group difference between the rotational velocities fitted for each group minus the differential velocity law fitted with de la Rue's data, with the averages taken over equal width bins of area on a logarithmic scale. The distribution shows behaviour that is constant with area. The mean value is $-0.04 \pm 0.10$\ degrees/day.}
\label{Area_V}
\end{figure}

\citeauthor{1966Ward} (\citeyear{1966Ward}) and
\citeauthor{1984HowardGilmanGilman}
(\citeyear{1984HowardGilmanGilman}) studied the relationship
between the rotational velocity of a sunspot group and its
maximum area. We analysed this correlation in the data of
\citeauthor{1870delaRue} (\citeyear{1869delaRue},
\citeyear{1870delaRue}).  We did not consider the absolute
rotational velocity because of its dependence on the
heliographic latitude. Instead, we considered the difference
between the rotational velocity evaluated for each sunspot
group and the differential velocity calculated using the
coefficients given in Table \ref{omega_diff}. To this end,
we took a logarithmic scale for the areas and split them
into twelve equal width bins on that scale, and then
averaged the velocities in each of those bins. The results
are shown in Figure \ref{Area_V}.  One observes in the
figure that the differences do not depend on the area,
the mean value being $\Delta \omega = -0.04 \pm 0.10$\
degrees/day.

\section{Conclusions}
\label{conclusions}
Solar rotation is an important parameter in trying to
understand our Sun.  There is little information available
on solar rotation in the 19th century, however.  We have
tried to contribute to improving this situation by studying
the sunspot catalogues published by Carrington, Peters, and de la
Rue in that century, covering the period 1853--1870. We
have determined the rotational velocity for each sunspot
group identified in those catalogues. From our
findings, we would highlight three results:

(i) We found no differences in rotational
velocity between sunspot groups that were observed
simultaneously by two observers.

(ii) Fitting a differential rotation law $(\omega = a +
b\cdot sin^2B)$, we found no differences between Peters'
and de la Rue's observations. There was, however, a slightly lower value
for $a$, the rotational velocity at the equator, for
Carrington's observations.  There were no
differences between the values of $b$, the coefficient of the
term that depends on the latitude, considering the error
bars.

(iii) Finally, we found no differences in the rotational
velocity between small groups and large groups using the
sunspot data published by \citeauthor{1870delaRue}
(\citeyear{1869delaRue}, \citeyear{1870delaRue}).

This work was made possible by the recovery of three sunspot
catalogues published in the 19th century. The study and
merging of historical sunspot catalogues should contribute to
our better understanding the state of the Sun in the last two
centuries at least (\citealp{2014Lefevre};
\citealp{2014Carrasco}).

%
\begin{acks}
  Ricard Casas acknowledges financial support received from the Spanish Ministerio de Ciencia e Innovaci\'on (MICINN), project AYA2012-39620, Consolider-Ingenio CSD2007-00060, and research project 2009SGR1398 from the Generalitat de Catalu\-nya.  Jos\'e M.  Vaquero acknowledges financial support from the Junta de Extremadura (Research Group Grant No. GR10131), from the Ministerio de Economia y Competitividad of the Spanish Government (AYA2011-25945), and from the COST Action ES1005 TOSCA\footnote{http://www.tosca-cost.eu}.

\end{acks}

%
%
%

\end{article} 
\end{document}